# Nanoscale Probing of Broken-Symmetry States in Graphene Induced by Individual Atomic Impurities

Yu Zhang[1], Qi-Qi Guo[1], Si-Yu Li[1], and Lin He[1,2,*]

[1] Center for Advanced Quantum Studies, Department of Physics, Beijing Normal University, Beijing, 100875, People's Republic of China.

[2] State Key Laboratory of Functional Materials for Informatics, Shanghai Institute of Microsystem and Information Technology, Chinese Academy of Sciences, 865 Changning Road, Shanghai 200050, People's Republic of China

[*]Correspondence and requests for materials should be addressed to L.H. (e-mail: helin@bnu.edu.cn).

**Inherent symmetries of a system lead to multiple degeneracies of its energy spectra. Introducing individual atomic impurities can locally break these symmetries, which is expected to lift the degenerate degrees of freedom around the impurities. Although central to our understanding of the fundamental properties of solids, the broken-symmetry states induced by individual atomic impurities have so far eluded observation. Here, we report nanoscale probing of the broken-symmetry states in graphene induced by two types of individual atomic impurities, i.e., isolated nitrogen dopants and isolated hydrogen atoms chemisorbed on graphene. Our experiments demonstrate that both types of atomic impurities can locally break sublattice symmetry of graphene and generate valley-polarized states, which extends several nanometers around the impurities. For the isolated hydrogen atom chemisorbed on graphene, the enhanced spin-orbit coupling, which arises from the $sp^3$ distortion of graphene due to the hydrogen chemisorption, further lifts the spin degeneracy, resulting in a fully spin and valley polarized states within about 1 nm around the hydrogen atom. Our result paves the way to control various broken-symmetry states at the nanoscale by various atomic impurities.**

In solids, such as graphene, with multiple degenerate degrees of freedom in energy spectra, it is possible to realize many exotic broken-symmetry states[1-8]. It is well-known that the fourfold valley and spin degeneracies in graphene arise through its inherent symmetries. In previous studies, the degeneracy of graphene is globally lifted, for example, by electron-electron interactions, and then the exotic ordered states, such as multicomponent quantum Hall states, ferromagnetism, and even superconductivity, are realized in the whole system[9-18]. Introducing individual atomic impurities could locally break the inherent symmetries of graphene[19-33] and, therefore, it is expected to lift the degeneracies of the energy spectra and realize novel broken-symmetry states in graphene at the nanoscale. However, measuring broken-symmetry states induced by an isolated atomic impurity in graphene has turned out to be challenging.

In this Letter, two types of individual atomic impurities, i.e., isolated nitrogen dopant and chemisorbed single hydrogen atom, are introduced in a controlled manner in graphene. By using Landau levels (LLs) spectroscopy, we systematically measure the broken-symmetry states induced by the two types of individual atomic impurities. Our experiment demonstrate that both the nitrogen dopant and hydrogen atom can lift the valley degree of freedom for several nanometers around the impurities by breaking sublattice symmetery of graphene. For the chemisorbed single hydrogen atom, a fully valley and spin polarized state is realized within about 1 nm of the impurity because the coexistence of sublattice symmetry breaking and enhanced spin-orbit coupling (SOC) introduced by the H atom. Our result highlights the way to tailor various broken-symmetry states in graphene at the nanoscale.

In our experiment, two types of individual atomic impurities, *i.e.*, nitrogen (N) dopant with a planar configuration[19] and hydrogen (H) chemisorption with an out-of-plane configuration[20], are introduced in a controlled manner in graphene, as schematically shown in Fig. 1a. The N dopants in graphene are introduced by thermal decomposition of a small amount of ammonia borane during the growth process of graphene[23]. The isolated H atoms absorbed on graphene is introduced via a $H_2$ plasma after the synthesis of graphene[34] (see Fig. S1-S3 and Methods of the Supplemental Material[35] for details). The obtained impurities are of extremely low concentrations and, usually, there is only one atomic impurity within a 20 × 20 nm$^2$ region to remove any possible interactions between the impurities. To explore the degeneracy and broken-symmetry states around the individual atomic impurities, we measure Landau levels (LLs) spectra around the impurities by carrying out scanning tunnelling microscopy and spectroscopy (STM and STS) measurements in high magnetic fields. In pristine graphene monolayer, a perpendicular magnetic field quantizes the continuous electronic spectrum into discrete LLs, and each LL is fourfold degenerate due to spin and valley degrees of freedom[36-39], as schematically shown in Fig. 1b. When the degeneracy of graphene is partially lifted, for example, the valley degeneracy is lifted, each LL will split into two peaks (Fig. 1c). If both the valley and spin degrees of freedom in graphene are removed, each LL will split into four peaks, as schematically shown in Fig. 1d. By taking advantage of high spatial resolution of the STM and high energy resolution of the LLs spectra, as shown subsequently, we can measure the subtle broken-symmetry states and their spatial extention around the individual atomic impurities of graphene.

Figure 2a shows a representative STM image of graphene with an individual N atom substituting for a C atom and the inset of Fig. 2a shows the zoom-in image of the N dopant. A triangle-like topographic feature with the maximum apparent out-of-plane height ~ 85 pm can be clearly identified (see Fig. S4 of the Supplemental Material[35]). The STS spectrum recorded on the N dopant (Fig. S4) exhibits a prominent electron-hole asymmetry, accompanied by a resonant peak at ~ 0.5 eV above the Fermi energy. All these features, including the STM image and the STS spectrum, are well consistent with that of an isolated N dopant in graphene, as reported in previous studies[19,40,41]. The spatial-resolved broken-symmetry states of graphene around the N dopant are studied via the LL spectra under high magnetic fields. The STS spectrum recorded at 5 nm away from the N atom exhibits a well-defined Landau quantization of the massless Dirac fermions (Fig. 2b), which demonstrates explicitly that the topmost graphene monolayer efficiently decouples from the supporting substrate and behaves as a free-standing graphene monolayer[42-44]. Moreover, no detectable splitting of the LLs is observed, indicating that the fourfold degeneracy of the graphene monolayer is preserved. However, for the STS spectrum recorded at the N atom, a notable splitting of the $N = 0$ LL, ~ 20 meV, is clearly observed (Fig. 2b).

Figure 2c summarizes the splitting of the 0 LL as a function of distance $L$ away from the N atom under $B = 8$ T. At the positions with the distance $L$ that is larger than 3.8 nm, no splitting is detectable. With the measured position approaching the N atom, the 0 LL gradually splits into two peaks and the splitting reaches the maximum when recorded on the N atom, as shown in Fig. 2c (similar feature has also been observed in other different magnetic fields, see Fig. S5[35]). Such a result indicates explicitly that the splitting of the 0 LL is induced by the N dopant. For graphene with sublattice symmetry breaking $\Delta$, the energies of the 0 LL in the K and K′ valleys will be shifted in opposite directions and consequently the 0 LL will be split into two peaks[39,45,46]. The N dopant locally breaks the sublattice symmetry of graphene and it is expected to generate valley-polarized 0 LL around the impurity. Therefore, the splitting of the 0 LL observed in our experiment is attributed to the realization of valley-polarized state induced by the individual N dopant. Such a result is further confirmed by magnetic field dependence of the splitting measured at the N atom, as shown in Fig. 2d. The magnetic fields further enlarge the energy separations of the splitting, yielding a linear relation with the slope of 2.5 meV/T. Both the linear relation of the energy separations as a function of magnetic fields and the magnitude of the slope are consistent with the valley splitting observed in graphene monolayer[47,48]. In Fig. 2e, we show the spatial region around the N dopant where we can detect the valley splitting of the 0 LL as a function of magnetic fields. Obviously, the valley-polarized state induced by the individual N dopant can extend several nanometers around the impurity.

For individual H atoms chemisorbed on graphene, our experiment demonstrates that even richer broken-symmetry states are introduced around the impurity. Figure 3(a) shows a typical STM image of an individual H atom chemisorbed on the topmost graphene monolayer. A bright protrusion with a lateral height up to ~ 200 pm, surrounded by a triangular $(\sqrt{3} \times \sqrt{3}) R30°$ interference pattern, can be clearly identified (Fig. S6 of the Supplemental Material[35]). The STS spectrum recorded on the H atom shows a quasi-localized $V_\pi$ state at ~ 0.1 eV above the Fermi level (Fig. S6 of the Supplemental Material[35]). The features of the STM image and the existence of the localized $V_\pi$ state in the STS spectrum are well consistent with that of an isolated H atom chemisorbed on graphene[20]. The chemisorbed H atom, which is similar as the N dopant, also locally breaks the sublattice symmetry of graphene. Therefore, we also detect the valley polarized state, as featured by splitting into two peaks of the 0 LL, around the H atom (Fig. 2b) and the valley splitting also increases linear with external magnetic fields (see Fig. S7 of the Supplemental Material[35]). However, when we measure the LLs spectrum at the chemisorbed H atom, two new features are observed, as shown in Fig. 2b. First, each of the valley-polarized peak of the 0 LL further splits into two peaks, indicating that the spin degeneracy is lifted[9,47]. Second, we observe splitting of nonzero LLs: for example, the -1 LL clearly splits into two peaks. The spin splitting of the K' and K valleys in the 0 LL, as defined as $\Delta E_1^{0\ LL}$ and $\Delta E_2^{0\ LL}$, when recorded on the H atom depend weakly on the magnetic fields, as shown in Fig. 3(c). The splitting of the nonzero LL is also measured in different magnetic fields and Fig. 3(d) shows the splitting of the -1 LL as a function of magnetic fields as an example (see Fig. S8 of the Supplemental Material for splitting of the other nonzero LLs[35]). The observed fully spin and valley polarized states of the 0 LL and the splitting of the nonzero LLs are attributed to the coexistence of sublattice symmetry breaking and enhanced SOC due to the chemisorption of the H atom. According to previous studies, the hybridization of a H atom and the underlying graphene C atom could induce a distortion of the graphene lattice from *sp²* to *sp³*, which leads to a significant enhancement of the SOC in graphene around the H impurity[49,50].

To further understand the nature of the broken-symmetry states induced by the chemisorption of the H atom, we carry out the tight-binding calculations of the subtle energies in the LLs of graphene under perpendicular magnetic fields[51-53] (see Supplemental Material for details[35]). By considering both the sublattice symmetry breaking and the enhanced SOC, the fourfold spin-valley degeneracy of the zero LL of graphene is completely lifted. The energies of the four spin-valley polarized states can be described as

$$E_{K\downarrow} = \Delta + \lambda_A, \ E_{K\uparrow} = \Delta - \lambda_A, \ E_{K'\downarrow} = -\Delta + \lambda_B, \ E_{K'\uparrow} = -\Delta - \lambda_B.$$

Here $\Delta$ is potential difference between the nearest carbon site induced by the H adatom, $\lambda_A$ and $\lambda_B$ are the strength of atomic SOC on the H absorbed carbon site and the nearest carbon sites, respectively [Fig. 3(e) and Tab. S1 of the Supplemental Material[35]]. Because the sublattice potential asymmetry induced by the H adatom is larger than the strength of the atomic SOC, the experimental obtained $\Delta E_1^{0\ LL}$ and $\Delta E_2^{0\ LL}$ can roughly reflect the strength of the atomic SOC induced by the H adatom, yielding $\lambda_A = 4.5\ meV$ and $\lambda_B = 2.0\ meV$. Theoretically, the $sp^3$ distortion induced by the H adatom can efficiently enhances the localized interaction between $\pi$ and $\sigma$ bands, resulting in an enhanced atomic SOC, which is predicted to be about 7 meV, acting on the $\pi$ electrons of graphene[49] (Fig. S9 of the Supplemental Material[35]). Experimentally, the SOC in weakly hydrogenated graphene with 0.05% H absorptions is measured as 2.5 meV in transport experiment, which is consistent with that obtained in our experiment. The coexistence of the sublattice symmetry breaking and the enhanced atomic SOC around the H adatom not only lead to the fully spin and valley polarized zero LL, but also lift the valley degeneracy of the nonzero LLs, as shown in Fig. 3e. Additionally, the calculated splitting of the *N* < 0 LLs is slightly larger than that of the *N* > 0 LLs (Fig. 3e). Obviously, all the experimental features observed in our experiment are reproduced quite well by our theoretical calculation.

As demonstrated in Fig. 2, we can obtain the spatial extention of the sublattice symmetry breaking induced by the atomic impurity by measuring the valley splitting of the 0 LL. Similarly, we also can measure the spatial extention of the SOC induced by the chemisorbed H atom by measuring the fully spin and valley polarized 0 LL and the valley polarized nonzero LLs. In Fig. 4a, we plot the energy separations of valley splitting in both the 0 LL and the -1 LL as a function of distance away from the H atom under the magnetic field of 12 T. Three distinct regions, as marked by different colors in the top panel of Fig. 4a, can be clearly identified. In the nearest region of the H atom, within about 1 nm from the H atom, both the enhanced SOC and the sublattice symmetry breaking play important roles in determining the broken-symmetry states and we observe fully spin and valley polarized states of the zero LL and valley-polarized state of the nonzero LL. The spatial extention of the valley splitting in the -1 LL is slightly smaller than that of the fully spin and valley splitting of the zero LL, attributing to the decrease of the lifetime of quasiparticles with increasing the LL index (the spatial extention of the valley splitting decreases for higher LL indices, see Fig. S10 of the Supplemental Material[35]). In the region far away from the H atom, the effect of both the SOC and the sublattice symmetry breaking induced by the impurity can be neglected and we observe well-defined LLs with fourfold degeneracy (Fig. 4a and 4b). In the intermediate region, only the sublattice symmetry breaking dominates the breaking-symmetry states and we observe valley-polarized zero LL in our experiment. Figure 4b summarizes the three regions as a function of magnetic fields measured in our experiment. Obviously, the spatial extention of the SOC is much smaller than that of the sublattice symmetry breaking induced by the chemisorbed H atom.

In summary, we systematically study the broken-symmetry states in graphene induced by individual atomic impurities via the LL spectra. The broken-symmetry states induced by a chemisorbed H atom are quite different from that induced by a N dopant, indicating the possibility to tailor the broken-symmetry states of graphene at the nanoscale. This opens a road to realize and tailor more exotic broken-symmetry states in graphene by using different atomic impurities, such as heavy elements with a larger atomic SOC or magnetic elements.

**Acknowledgements**

This work was supported by the National Natural Science Foundation of China (Grant Nos. 11974050, 11674029). L.H. also acknowledges support from the National Program for Support of Top-notch Young Professionals, support from "the Fundamental Research Funds for the Central Universities", and support from "Chang Jiang Scholars Program".

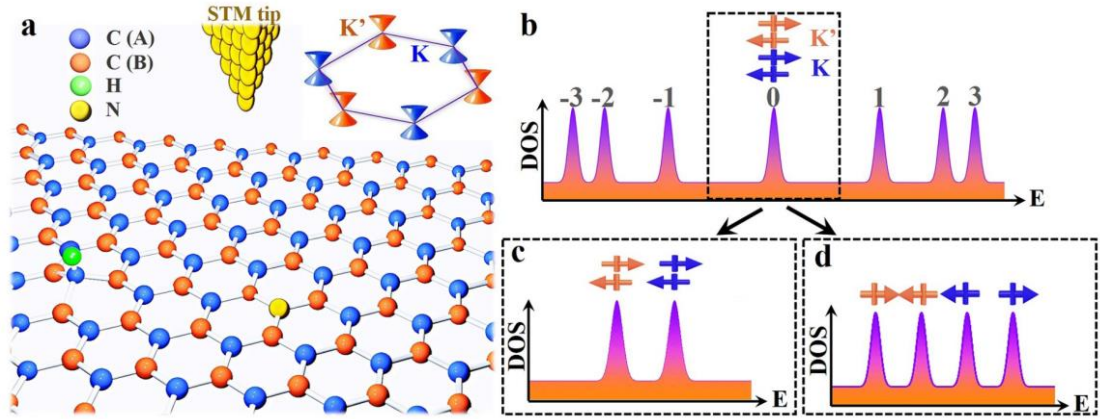

**Figure 1. Illustration of the broken-symmetry states induced by an individual atomic impurity in graphene monolayer.** (**a**) Schematic STM measurements of graphene monolayer with two types of individual atomic defects, i.e. nitrogen dopant and chemisorbed hydrogen atom. The inset is the schematic low-energy spectrum of graphene near the Dirac points in the first Brillouin zone, with two inequivalent valleys labeled as K and K'. (**b**) A schematic DOS of LLs in pristine graphene monolayer under high magnetic fields. The LL indices are labeled, and the $N = 0$ LL exhibits fourfold valley and spin degeneracies. (**c**) A schematic DOS of a valley polarized $N = 0$ LL with two valley-polarized peaks. (**d**) A schematic DOS of a fully spin and valley polarized $N = 0$ LL, which splits into four separate peaks.

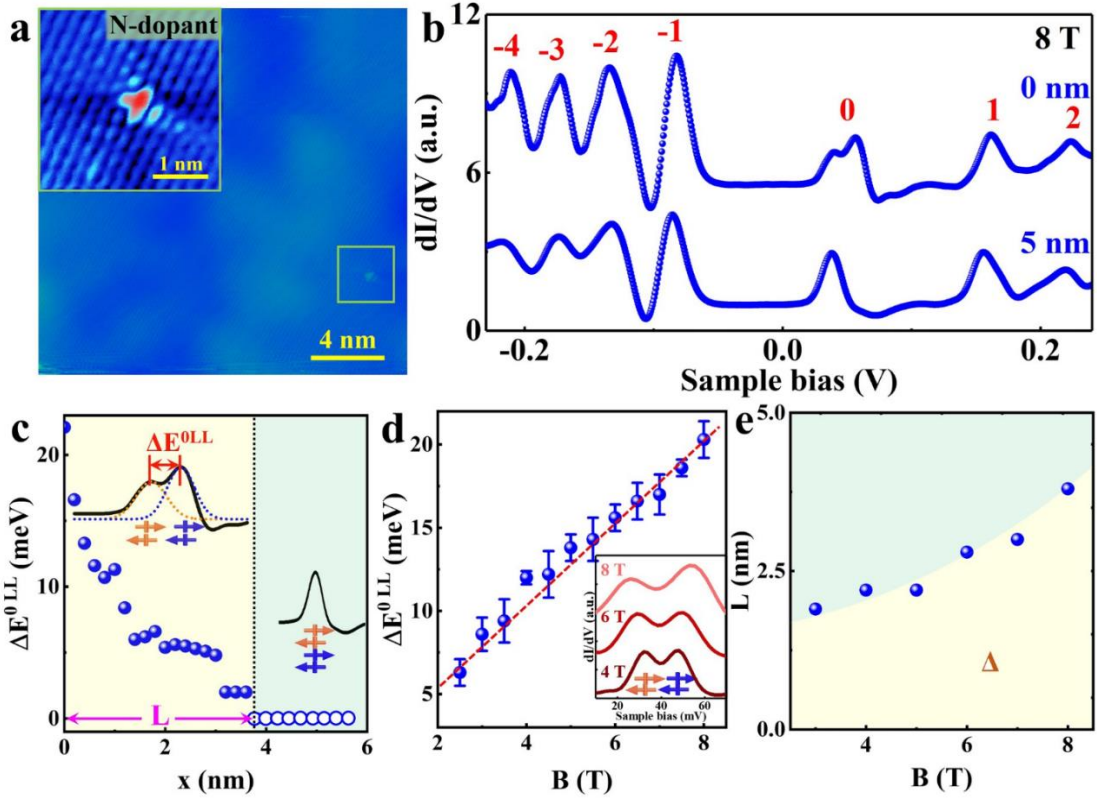

**Figure 2. Atomic-resolution STM images and STS spectra of an individual N dopant in graphene.** (a) A 20 × 20 nm² STM topography of the topmost graphene monolayer with an individual N dopant. Inset: the atomic-resolution STM image of an individual N dopant in graphene. (b) STS spectra measured on the N atom and at a position ~ 5 nm away from the N atom under the magnetic field of 8 T. The spectra are offset vertically for clarity. The LL indices are labeled, and the splitting of the $N = 0$ LL recorded on the N atom can be clearly identified. (c) Energy separations of the K and K' valley in the $N = 0$ LL as a function of the distance from N site under the magnetic field of 8 T. At the position with the distance L away from the N atom, the splitting decreases to zero. (d) Energy separations of the K and K' valley in the $N = 0$ LL under different magnetic fields recorded on the N site. Inset: Typical STS spectra of the $N = 0$ LL recorded on the N site under different magnetic fields. (e) The distance L, as defined in panel d, as a function of magnetic fields. All the experimental data were acquired at $V_{bias} = 0.5\ V, I = 0.2\ nA$.

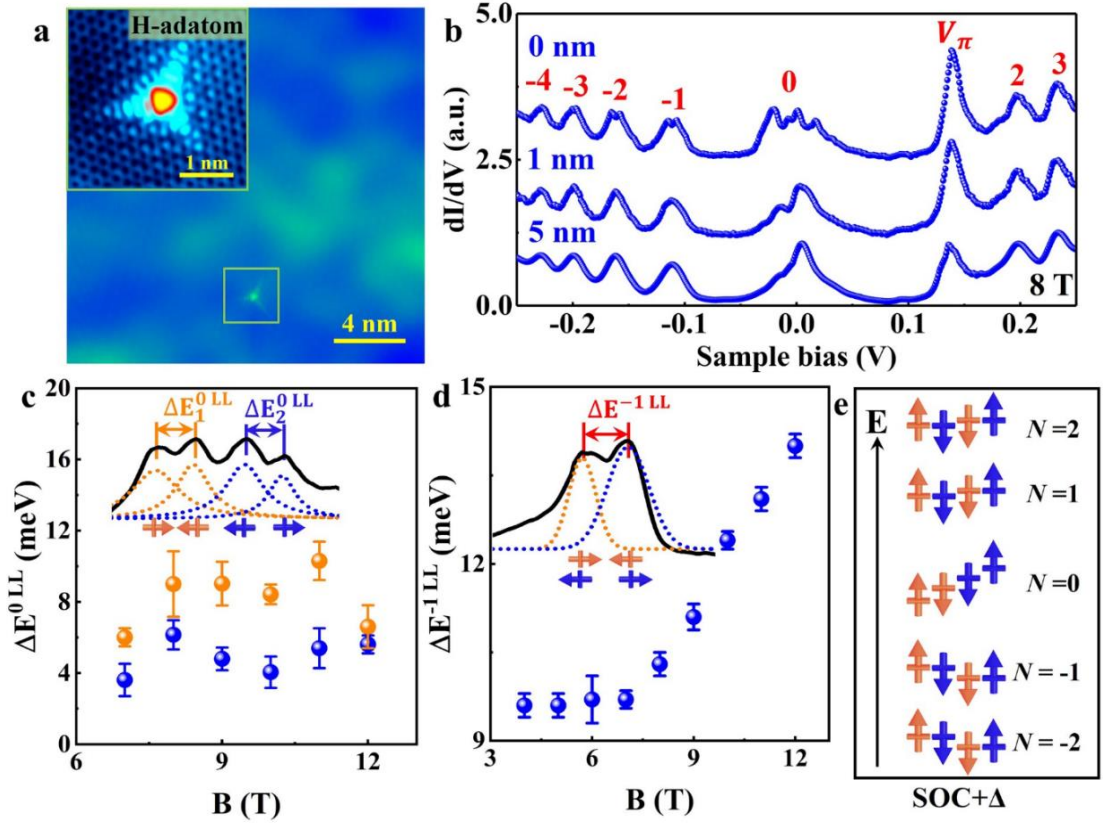

**Figure 3. Atomic-resolution STM images and STS spectra of an individual H chemisorbed on graphene.** (**a**) A 20 × 20 nm² STM topography of the topmost graphene monolayer with an individual H atom. Inset: the atomic-resolution STM image of the chemisorbed H atom. (**b**) STS spectra recorded at different distances away from the H adatom under the magnetic field of 8 T. The spectra are offset vertically for clarity. (**c**) Energy separations of spin splitting in each valley of the $N = 0$ LL recorded on the H site under different magnetic fields. The $\Delta E_1^{0\,LL}$ and $\Delta E_2^{0\,LL}$ are defined as the spin splitting in each valley, as shown in the inset. (**d**) Energy separations of the $N = -1$ LL recorded on the H site as a function of magnetic fields. $\Delta E^{-1\,LL}$ is defined as the splitting of the -1 LL, as shown in the inset. All experimental data were acquired at $V_{bias} = 0.5\,V, I = 0.2\,nA$. (**e**) Schematic image for the electron energies in different LLs with considering the sublattice symmetry breaking and the enhanced SOC in graphene.

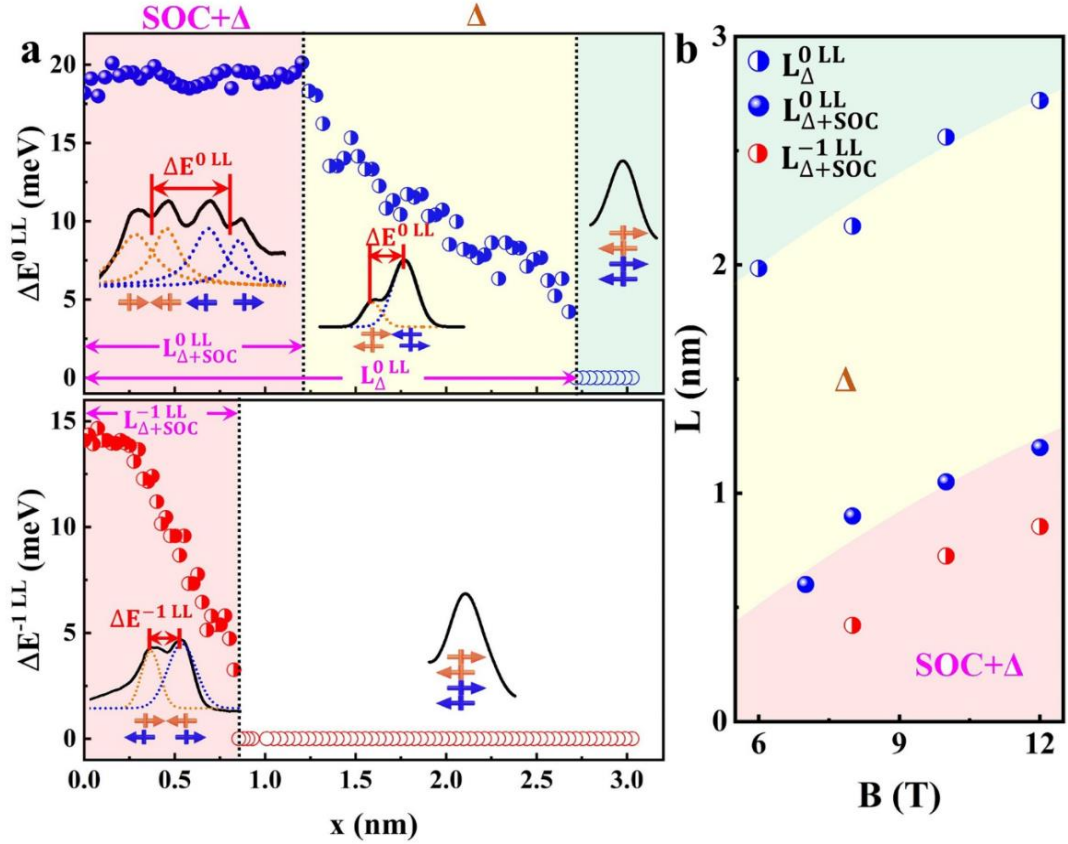

**Figure 4. The spatial distributions of the splitting of the LLs around the chemisorbed H atom on graphene.** (**a**) Up panel: The energy splitting of the K and K' valley in the $N = 0$ LL around the H atom on graphene as a function of the distance from H site under the magnetic fields of 12 T. Three different regions are marked by pink (the 0 LL splits into four peaks), yellow (the 0 LL splits into two peaks) and green shadow (there is no observable splitting of the 0 LL) respectively. The length $L^{0\,LL}_{\Delta+SOC}$ is defined as the distance from the H atom that the coexistence of sublattice symmetry breaking and SOC lead to the fully spin and valley splitting of the 0 LL. The $L^{0\,LL}_{\Delta}$ is defined as the distance from the H atom that the valley splitting of the 0 LL is observable. Bottom panel: The energy splitting of the $N = -1$ LL around the chemisorbed H atom on graphene as a function of the distance from H site under the magnetic fields of 12 T. The length $L^{-1\,LL}_{\Delta+SOC}$ is defined as the distance from the H atom that the coexistence of sublattice symmetry breaking and SOC lead to the valley splitting of the -1 LL. (**b**) The measured lengths $L^{0\,LL}_{\Delta+SOC}$, $L^{0\,LL}_{\Delta}$ and $L^{-1\,LL}_{\Delta+SOC}$ as a function of magnetic fields.